# ReStore: Reusing Results of MapReduce Jobs

Iman Elghandour
University of Waterloo
ielghand@cs.uwaterloo.ca

Ashraf Aboulnaga
University of Waterloo
ashraf@cs.uwaterloo.ca

## ABSTRACT

Analyzing large scale data has emerged as an important activity for many organizations in the past few years. This large scale data analysis is facilitated by the MapReduce programming and execution model and its implementations, most notably Hadoop. Users of MapReduce often have analysis tasks that are too complex to express as individual MapReduce jobs. Instead, they use high-level query languages such as Pig, Hive, or Jaql to express their complex tasks. The compilers of these languages translate queries into workflows of MapReduce jobs. Each job in these workflows reads its input from the distributed file system used by the MapReduce system and produces output that is stored in this distributed file system and read as input by the next job in the workflow. The current practice is to delete these intermediate results from the distributed file system at the end of executing the workflow. One way to improve the performance of workflows of MapReduce jobs is to keep these intermediate results and reuse them for future workflows submitted to the system. In this paper, we present *ReStore*, a system that manages the storage and reuse of such intermediate results. ReStore can reuse the output of whole MapReduce jobs that are part of a workflow, and it can also create additional reuse opportunities by materializing and storing the output of query execution operators that are executed within a MapReduce job. We have implemented ReStore as an extension to the Pig dataflow system on top of Hadoop, and we experimentally demonstrate significant speedups on queries from the PigMix benchmark.

## 1. INTRODUCTION

Massive scale data analysis has become a main activity for many enterprises and research groups. Companies such as Facebook, Yahoo, and Google now own petabyte-scale data warehouses that are accessed on a regular basis using ad hoc queries and periodic batch jobs [7, 16], and terabyte-scale data warehouses are now common in many smaller companies. This large scale data analysis is currently supported by the MapReduce programming and execution model [9] and its implementations such as Hadoop [1], which is now one of the major platforms for data analysis.

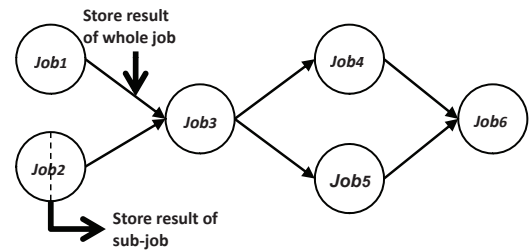

**Figure 1: An example workflow of MapReduce jobs.**

Users of MapReduce often have analysis tasks that are too complex to express as one MapReduce job. Instead, they often use high-level query languages such as Pig Latin [11, 14], Hive [15], or Jaql [8] to express their complex analysis tasks. The compilers of these query languages translate queries into *workflows of MapReduce jobs*, such as the one shown in Figure 1. Optimizing the performance of such workflows is important given the popularity of MapReduce and these query languages on top of it.

Each job in a workflow of MapReduce jobs produces output that is stored in the distributed file system used by the MapReduce system (e.g., HDFS [3] in the case of Hadoop). These intermediate results are used as input by subsequent jobs in the workflow. For example, *Job3* in Figure 1 produces output that is used as input by *Job4* and *Job5*. The current practice is to delete these intermediate outputs after finishing the execution of the workflow. In this paper, we present *ReStore*, a system that improves the performance of workflows of MapReduce jobs generated from high-level query languages by storing the intermediate results of executed workflows and reusing them for future workflows submitted to the system.

We expect reusing the output of MapReduce jobs to be beneficial since it is common for enterprises to have large data sets on which many data analysis queries are executed (e.g., the usage log data in internet companies such as Facebook). Queries on these data sets typically perform the following steps: (1) load the data set, (2) perform some simple processing to filter out unnecessary data, and (3) perform extra processing on the small fraction of the loaded data that passes the filter. Steps 1 and 2 of one workflow are likely to be repeated in other workflows, and even parts of Step 3. These steps are repeated in many queries, and the MapRe-duce jobs that execute them can be replaced by reading the stored outputs of similar jobs that were executed in past workflows, and whose output we have stored using ReStore. Moreover, even if full jobs cannot be reused, parts of a job (which we call a *sub-job*) can be useful for future workflows, and ReStore can materialize and store the outputs of such sub-jobs.

Finding sharing opportunities among queries that are submitted in the same batch to a MapReduce cluster has been studied in [5] and [13], but these works focus on sharing between queries that are executed concurrently and are limited to sharing one operator between multiple queries. In this paper, we enable queries submitted at different times to share results and we can share large portions of the executed workflows. The importance of sharing is illustrated by the fact that Facebook stores the result of any query in its MapReduce cluster for seven days so that it can be shared among users [16].

ReStore can be built on top of dataflow language processors such as Pig, Hive, or Jaql. These language processors translate queries into workflows of MapReduce jobs. Each of these MapReduce jobs has a *physical query execution plan* that contains one or more *physical operators* that are executed by this job. Each language has a fixed set of physical operators such as *Filter*, *Select*, and *Join*. The workflows of MapReduce jobs are submitted to ReStore, which performs the following: (1) it rewrites the MapReduce jobs in a submitted workflow to reuse job outputs previously stored in the system, (2) it stores the outputs of executed jobs for future reuse, (3) it creates more reuse opportunities by storing the outputs of sub-jobs in addition to whole MapReduce jobs, and (4) it selects the outputs of jobs to keep in the distributed file system and those to delete. After ReStore rewrites a MapReduce job, it submits it to the MapReduce system to be executed. These steps can be viewed as analogous to the steps of building and using materialized views for relational databases [12]. In this paper, we focus on describing steps 1–3 and present a brief discussion of techniques for performing step 4.

In the rest of the paper we present the following contributions:

- A framework for creating reuse opportunities between workflows of MapReduce jobs and taking advantage of these opportunities (Section 2).

- A technique to rewrite input workflows of MapReduce jobs to reuse the results of previously executed jobs stored in the system (Section 3).

- A technique to increase reuse opportunities by materializing the outputs of sub-jobs of the executed MapReduce jobs (Section 4).

- A proposal of a primitive set of heuristic rules for deciding which of the candidate MapReduce job outputs to keep and which to discard (Section 5).

- An implementation of ReStore on top of the Pig system [11, 14] (Section 6), and an experimental study using this implementation (Section 7).

## 2. OVERVIEW OF RESTORE

The compilers of dataflow languages such as Pig Latin or Hive translate an input SQL-like query into a physical query execution plan that consists of physical operators such as *Filter*, *Select*, and *Join*. The compiler then embeds the operators of this query execution plan into a workflow of MapReduce jobs. The reason for having a workflow of MapReduce jobs and not just one MapReduce job is that some physical operators such as *Join* and *Group* need to be divided between a mapper stage and a reducer stage [14]. Consequently, when more than one of these physical operators exist in a query execution plan, each one of them has to be embedded in a separate MapReduce job. Each generated MapReduce job reads its inputs from the distributed file system using one or more *Load* operators and stores its output in the distributed file system using a *Store* operator. After embedding all physical operators into mapper and reducer stages, the result is a workflow of MapReduce jobs, each with its own physical query execution plan. The compiler then generates code for each MapReduce job in the workflow and passes this job to the MapReduce system (e.g., Hadoop) for execution. ReStore extends such a dataflow system by adding functionality that enables future workflows to reuse the output of full MapReduce jobs with all the physical operators that they contain, or the output of sub-jobs representing some of the physical operators within a MapReduce job.

To illustrate the various result reuse opportunities in this paper, we use as an example two queries Q1 and Q2[1]. The workflows of MapReduce jobs for these queries and the physical query execution plans that are embedded in these jobs are shown in Figures 2 and 3. These workflows would be the inputs to ReStore. The Q2 physical plan in Figure 3 is divided into two MapReduce jobs because each of the *Join* and *Group* operators in Q2 needs to be placed in a separate reducer stage.

**Query Q1 (based on PigMix L2): Return the estimated revenue for each user viewing web pages**

```
A = load 'page_views' as (user, timestamp,
          est_revenue, page_info, page_links);
B = foreach A generate user, est_revenue;
alpha = load 'users' using (name, phone,
                            address, city);
beta = foreach alpha generate name;
C = join beta by name, A by user;
store C into 'L2_out';
```

**Query Q2 (based on PigMix L3): Return the total estimated revenue for each user viewing web pages, grouped by user name**

```
A = load 'page_views' as (user, timestamp,
          est_revenue, page_info, page_links);
B = foreach A generate user, est_revenue;
alpha = load 'users' using (name, phone,
                            address, city);
beta = foreach alpha generate name;
C = join beta by name, A by user;
D = group C by $0;
E = foreach D generate group, SUM(C.est_revenue);
store E into 'L3_out';
```

Next, we describe the reuse opportunities that we exploit in this paper and the expected benefit of such reuse, and in Section 2.2 we present the ReStore architecture.

---
[1]Throughout this paper, we use queries that are written in Pig Latin [14] and that come from the PigMix benchmark [4].

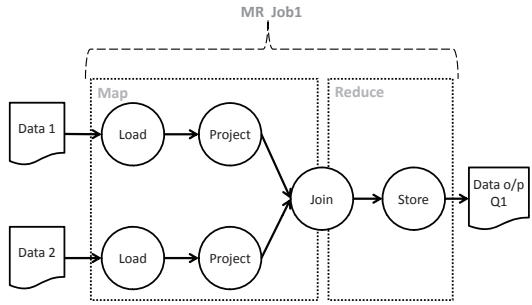

Figure 2: The MapReduce workflow for query Q1.

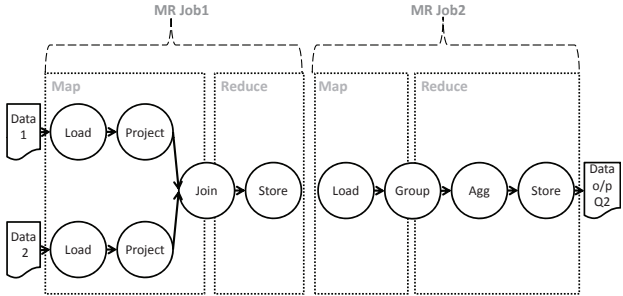

Figure 3: The MapReduce workflow for query Q2.

## 2.1 Types of Result Reuse in ReStore

Every job in the workflow does not start until all the jobs on which it depends finish. Therefore, the total time needed to execute $Job_n$, $T_{total}(Job_n)$, equals the time needed to finish executing the job, $ET(Job_n)$, in addition to the time needed to finish all jobs, $\forall_{i \in Y} Job_i$, where $Y$ is the set of jobs on which $Job_n$ depends, which is the time needed for the slowest of these jobs to finish. Hence, $T_{total}(Job_n)$ can be expressed as:

$$T_{total}(Job_n) = ET(Job_n) + max_{i \in Y}\{T_{total}(Job_i)\} \qquad (1)$$

ReStore generates two types of reuse opportunities by materializing the output of: (1) whole jobs, which reduces $max_{i \in Y}\{T_{total}(Job_i)\}$ in future workflows, and (2) operators in jobs (sub-jobs), which reduces $ET(Job_n)$ in future workflows. Next, we discuss these two reuse opportunities.

A MapReduce job can appear in multiple workflows that are generated by the dataflow language compiler for different queries that are submitted to the system at different times. Keeping the output of any job, $J_A$, in the distributed file system and reusing it in future workflows that have $J_A$ reoccurring in them is expected to reduce the execution time of these workflows. For example, Q1 (Figure 2) joins two data sets in a MapReduce job, and Q2 also joins the same two data sets and then performs grouping and aggregation on the result of the join. Thus, if we keep the output of Q1 in the distributed file system, we can rewrite Q2 to reuse this output. Figure 4 shows the rewritten workflow of Q2.

In Equation 1, if all dependant jobs of $Job_n$ were previously executed and are stored in the system, the total time to execute $Job_n$ can now be reduced to: $T_{total}(Job_n) = ET(Job_n)$. If a subset of these jobs, $X \subset Y$, is not already stored in the system and therefore we still need to execute these jobs before $Job_n$, the total time to execute $Job_n$ when rewritten to reuse stored job outputs is reduced only if $max_{i \in X}\{T_{total}(Job_i)\}$ is less than $max_{i \in Y}\{T_{total}(Job_i)\}$.

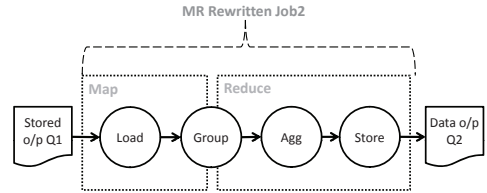

Figure 4: The MapReduce workflow for query Q2 after rewriting it to reuse the output of query Q1.

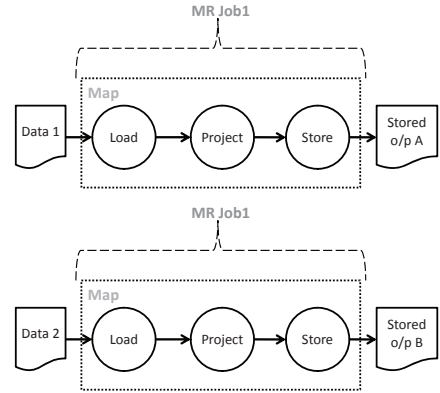

Figure 5: Different MapReduce jobs that can be used to rewrite query Q1.

Reusing the output of previously executed MapReduce jobs can significantly reduce the execution time of a workflow. However, it may not be easy to find a complete MapReduce job occurring unchanged in multiple workflows. It is more likely to find a *DAG of physical operators* that forms part of the query execution plan within one MapReduce job in one workflow occurring again in a MapReduce job in another workflow. The second type of reuse in ReStore is to materialize the output of such a DAG of physical operators that is part of a MapReduce job (i.e., a sub-job) and store it for future reuse. For example, consider the physical plan of Q1 shown in Figure 2. The plan starts with *Load* and *Project* operators that read data from two different data sources and discard the unnecessary columns. The output of the two *Project* operators is then pipelined into a *Join* operator. If we assume that the *Load* and *Project* operators on the two data sources form sub-jobs whose outputs were previously materialized and stored by ReStore, we can load these materialized outputs and pipeline them into the *Join* operator. Figure 5 shows the MapReduce jobs that produce these outputs, and Figure 6 shows Q1 after rewriting it to reuse these outputs.

The execution time of a job can be modeled as:

$$ET(Job_n) = T_{load} + \sum_i ET(OP_i) + T_{sort} + T_{store} \qquad (2)$$

where $T_{load}$ is the time required to load the data, $ET(OP_i)$ is the time required to execute physical operator $OP_i$, $T_{sort}$ is the time required to sort and shuffle the data between the mappers and reducers of the job, and $T_{store}$ is the time required to store the final output of the job. Reducing the size of the loaded data will reduce $T_{load}$. Furthermore, loading the stored outputs of some of the operators also reduces the time needed to execute these operators, $\sum_i ET(OP_i)$. These reductions add up and reduce $ET(Job_n)$.

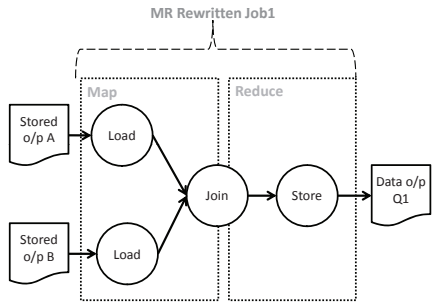

**Figure 6: The MapReduce workflow for query Q1 after rewriting it to reuse the outputs of sub-jobs.**

## 2.2 ReStore System Architecture

Figure 7 shows the main components of ReStore and how it is connected to the dataflow and MapReduce systems on which it is built. The input to ReStore is a workflow of MapReduce jobs generated by a dataflow system for an input query. The outputs are: (1) a modified MapReduce workflow that exploits prior jobs executed in the MapReduce system and stored by ReStore, and (2) a new set of job outputs to store in the distributed file system.

ReStore keeps a repository to manage the stored MapReduce job outputs. This repository contains for each stored job output: (1) the physical query execution plan of the MapReduce job that was executed to produce this output, (2) the filename of the output in the distributed file system, and (3) statistics about the MapReduce job that produced the output and the frequency of use of this output by different workflows. The physical plan of a MapReduce job contains information about the input data, the output data, and the operators that were executed to compute the output data. For example, we store in the repository the physical plan shown in Figure 2 along with the filename in the distributed file system of the output of the MapReduce job that executed Q1. We also store in the repository statistics about the MapReduce job that executed the physical plan, such as the size of the input, the size of the output, and the average execution time of mappers and reducers. In addition, we store statistics about how frequently the physical plan was used to rewrite queries submitted to ReStore. In Section 5, we describe how these statistics can be used to evaluate the benefit of keeping results in the repository.

ReStore has three main components: (1) plan matcher and rewriter, (2) sub-job enumerator, and (3) enumerated sub-job selector. For each job in the input workflow of MapReduce jobs, the plan matcher and rewriter searches the ReStore repository for the outputs of past jobs that can be used to answer all or part of this input job. The plan matcher and rewriter then rewrites the input job to reuse any outputs that it finds. The result of rewriting an input workflow of MapReduce jobs is a new workflow that likely performs less work than the input workflow and potentially has fewer jobs. The next step is to identify operators in the MapReduce jobs of the rewritten workflow that we ex-pect can be reused by future MapReduce workflows, which is done by the second and third components of ReStore. The second component is the sub-job enumerator, which enumerates for a given input MapReduce job the subsets of the physical operators within this job (DAGs of physical operators, or sub-jobs) that can be materialized and stored in the distributed file system. After executing the rewritten workflow in the MapReduce system, the enumerated sub-job selector – the third component of ReStore – examines the statistics collected for the MapReduce jobs that make up this workflow and selects which job outputs to keep in the repository and which to discard.

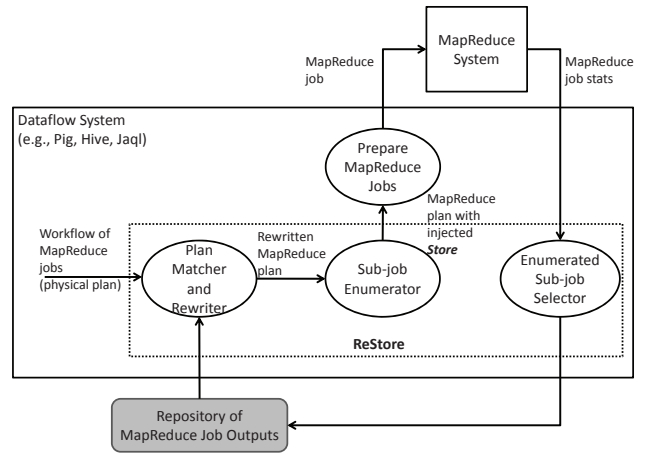

**Figure 7: ReStore system architecture.**

Note that matching, sub-job enumeration, and enumerated sub-job selection are based on physical plans. This makes ReStore portable across different dataflow systems since all these systems have similar physical operators. Customizing ReStore for a specific dataflow system only requires making it aware of the specific physical operators of the system. All algorithms would work with the new physical operators with little or no change. Next, we discuss the three components of ReStore in more detail.

## 3. MATCHING INPUT MAPREDUCE JOBS WITH PLANS FROM THE REPOSITORY

The first phase of ReStore is the plan matcher and rewriter. Matching and rewriting are performed on the physical plan of the input workflow of MapReduce jobs. The ReStore repository contains outputs of previously executed MapReduce jobs and the physical plans of these jobs, and the goal of matching and rewriting is to find physical plans in the repository that can be used to rewrite the jobs that make up the input workflow.

Matching and rewriting processes one MapReduce job at a time. Each MapReduce job in the input workflow is matched against the repository and rewritten to use job outputs in the repository if possible. The first jobs to be matched against the repository are the ones that read the input data sets, and the last jobs to be matched are the ones that produce the final results. Before a job $J$ is matched against the repository, all other jobs that $J$ depends on (i.e., the jobs whose outputs $J$ reads as input) have to be matched and rewritten to use the job outputs stored in the repository.

To match an input MapReduce job against the repository, ReStore scans sequentially through the physical plans in the repository and tests whether each plan matches the input MapReduce job. A physical plan in the repository is considered to match the input MapReduce job if this physical plan is contained within the physical plan of the input MapReduce job. As soon as a match is found, the input MapReduce job is rewritten to use the matched physical plan in the repository. Rewriting is done by identifying the part of the physical plan of the input MapReduce job that matches the physical plan selected from the repository. The matched part of the input physical plan is replaced with a *Load* operator that reads the output of the repository plan from the distributed file system. After rewriting, a new sequential scan through the repository is started to look for more matches to the rewritten MapReduce job. Thus, more than one physical plan in the repository can be used to rewrite an input MapReduce job. If a scan through the repository does not find any matches, ReStore proceeds to matching the next MapReduce job in the workflow.

It is possible that the physical plan in the repository matches a full MapReduce job $J$ in the input workflow (i.e., the entire physical plan for $J$ is already in the repository). In this case, other MapReduce jobs in the workflow that use the output of $J$ as input are rewritten so that they load their input data from the output of the repository plan instead of loading it from $J$. This enables the plan matcher and rewriter to use job outputs in the repository for all MapReduce jobs in the input workflow, even jobs whose input is the output of other jobs that are also stored in the repository.

Note that ReStore performs plan matching at the level of physical plans and not logical plans. The reason is that ReStore reuses results computed by past MapReduce jobs, and each of these MapReduce jobs is generated by the dataflow language compiler from a physical plan through a simple process of grouping physical operators into mappers and reducers. This direct correspondence between the physical plan and the MapReduce jobs makes matching simpler and more robust. Moreover, doing the matching at the physical plan level makes it easy to adapt ReStore to any dataflow system regardless of the input language and the logical query translation and optimization techniques used by the system. Adapting ReStore to a new system requires defining the physical operators of this system in ReStore, regardless of the optimization process that generates these operators.

Next, we turn our attention to the physical plan matching algorithm that is at the core of ReStore's plan matcher and rewriter. This algorithm tests whether a physical plan in the repository is contained in the physical plan of the input MapReduce job, and it is based on operator equivalence. Two operators are equivalent if: (1) their inputs are pipelined from operators that are equivalent or from the same data sets, and (2) they perform functions that produce the same output data. To match two plans, both plans are traversed simultaneously starting from the *Load* operators until mismatching operators are found or all the operators of the plan in the repository are found to have equivalent operators in the plan of the input MapReduce job.

Algorithm 1 illustrates the depth first traversal algorithm that we use to traverse two plans simultaneously. The *PairwisePlanTraversal* function is initially called with the *Load* operators of the input plan being examined by ReStore as *succsPlan1* and the *Load* operators of the plan from the

**Algorithm 1** *PairwisePlanTraversal(operator, succsPlan1, succsPlan2, seen, lastMatch)*

1: **if** $succsPlan2 == \phi$ **then**
2:     **return** *lastMatch*
3: **else if** $succsPlan1 == \phi$ **then**
4:     **return** null
5: **end if**
6: **for all** $succ \in succsPlan1$ **do**
7:     **if** $succ \notin seen$ **then**
8:       $seen \leftarrow seen \cup \{succ\}$
9:       $equivOP \leftarrow findEquivalentOP(succ, succsPlan2)$
10:      **if** $equivOP ==$ null **then**
11:        **continue** {to next *succ*}
12:      **else**
13:        $newSuccsPlan1 \leftarrow getSuccessors(succ)$
14:        $newSuccsPlan2 \leftarrow getSuccs(equivalentOP)$
15:        $retVal \leftarrow PairwisePlanTraversal(succ, newSuccsPlan1, newSuccsPlan2, seen, succ)$
16:        **if** $retVal ==$ null **then**
17:          **return** null
18:        **else**
19:          $succsPlan2 \leftarrow succsPlan2 - \{equivOP\}$
20:          **if** $succsPlan2 == \phi$ **then**
21:            **break**
22:          **end if**
23:        **end if**
24:      **end if**
25:     **end if**
26: **end for**
27: **return** *retVal*

repository as *succsPlan2*. In this initial call the parameter *seen* is set to the empty set and the parameter *lastMatch* is set to null. For every operator in the set of operators *succsPlan1*, we try to find an operator in *succsPlan2* that is equivalent to it (Line 9). If no equivalent operator is found, we continue trying to find an equivalent operator for other operators in *succsPlan1*. If an equivalent operator is found, we continue to match the successors of the operators that we found equivalent (Line 15). The successors of a physical operator are the operators that consume as one of their inputs the output produced by this operator. When the traversal of the successors of the operators *succ* and *equivOP* returns a value that is not null (Line 18), ReStore removes *equivOP* from the set of operators *succsPlan2* so it will not match with any other operators from *succsPlan1* (Line 19). If after the last step, *succsPlan2* becomes empty, we exit the loop (Line 21). If all the operators of the plan found in the repository have equivalent operators in the input plan being examined by ReStore, the repository plan is contained in the input plan and it is declared as a potential match for the input plan. As an example, Figures 4 and 6 show the workflows of MapReduce jobs for the queries in Figures 3 and 2, respectively, after matching and rewriting.

ReStore uses the first match that it finds in the repository to rewrite an input MapReduce job. This makes matching more efficient, but requires us to order the physical plans in the repository so that the first match found is the best match (i.e., the one that achieves the maximum reduction in the execution time of the input workflow). We use the following rules to order the physical plans in the repository:

1. Plan A is preferred to plan B if plan A subsumes plan B, meaning that all the operators in plan B have equivalent operators in plan A. For example, applying this rule on the plans in Figures 2 and 5, we can deduce that the former plan subsumes the latter plan and therefore ReStore should choose the former plan (Figure 2) to rewrite Q2 (Figure 3). This order of plans in the repository is enforced by the candidate generation algorithm described in Section 4.

2. If neither of plans A and B subsumes the other, we order them based on the following two metrics (the higher the better): (1) the ratio between the size of the input data and output data, and (2) the execution time of the MapReduce job. These metrics are calculated from the statistics collected by the MapReduce system during job execution after plan rewriting and sub-job generation.

## 4. GENERATING CANDIDATE SUB-JOBS FOR STORING IN THE REPOSITORY

Having described how to match the MapReduce jobs of a workflow with plans in the ReStore repository, we now turn our attention to populating the repository. In this section, we describe our approach for generating MapReduce job outputs that are candidates for storing in the repository. These candidate job outputs are all computed during the execution of the input MapReduce workflow (after it is rewritten to use plans in the repository) and stored in the distributed file system.

As discussed in Section 2.1, it is useful to store the output of whole MapReduce jobs and also the output of sub-jobs. Therefore, every MapReduce job output in ReStore is a candidate for including in the repository. The question is which sub-jobs should also be considered as candidates. We focus on this question in the rest of this section.

It is possible to treat the output of every physical operator $P$ in the physical plan of an input MapReduce job as a candidate sub-job. Let us call this sub-job $J_P$. This sub-job has a physical plan that contains all the physical operators in the input MapReduce job starting from the *Load* operators that read data from the distributed file system up to and including the operator $P$. If $P$, the last physical operator in $J_P$ is a *Store*, the output of $J_P$ would already be stored in the distributed file system after the execution of the input workflow. If $P$ is not a *Store*, we can add a *Store* as the last operator to ensure that the output of $J_P$ is stored in the distributed file system. Each candidate sub-job $J_P$ can be viewed as a complete MapReduce job that can be executed, stored, and matched independently of the rest of the input workflow. Sub-job $J_P$ is in fact stored in the repository as a full, independent MapReduce job that is indistinguishable from other jobs in the repository. As an example, Figure 5 shows two candidate sub-jobs that can be generated from the MapReduce job in Figure 2. Note that these two sub-jobs are *Map only* jobs.

Treating all possible sub-jobs as candidates and storing their outputs in the distributed file system during the execution of the input MapReduce workflow has two problems. First, it would require a substantial amount of storage in the distributed file system. Second, the overhead of storing all this intermediate data would considerably slow down the execution of the input MapReduce job. Furthermore, some of these sub-jobs may not be useful for future workflows. Thus, we need to select only a subset of the possible sub-jobs to consider as candidates, so that we can materialize and store these candidates in the distributed file system with reasonable cost. Some physical operators such as *Project* and *Filter* in Pig and similar systems are known to reduce the size of their input and are therefore good sub-job candidates according to Equation 2. Many MapReduce workflows are I/O bound [9], so reducing the size of the loaded data can significantly improve performance. Other physical operators such as *Join* and *Group* are known to be expensive, so their outputs are also good sub-job candidates because replacing them with stored output reduces $\sum_i ET(OP_i)$ in Equation 2.

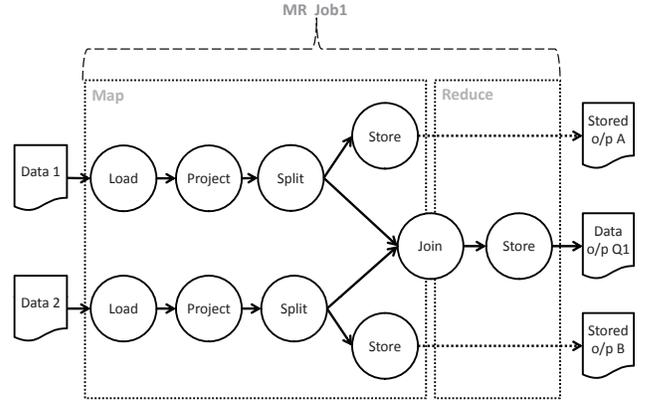

**Figure 8: The MapReduce workflow for query Q1 after introducing extra *Store* operators after *Project* operators.**

Based on this reasoning, it is possible to identify a set of physical operators that reduce their input size or are expensive, and to use the outputs of these operators as the candidate sub-jobs. This is indeed what we do in ReStore. To reduce the number of injected *Store* operators in the input physical plan, we propose the following two heuristics for choosing candidate sub-jobs:

1. **Conservative Heuristic:** Use the outputs of operators that are known to reduce their input size as candidate sub-jobs. These operators are *Project* and *Filter*.

2. **Aggressive Heuristic:** Use the outputs of operators that are known to reduce their input size and also the outputs of operators that are known to be expensive as candidate sub-jobs. These operators are *Project*, *Filter*, *Join*, *Group*, and *CoGroup*.

The conservative heuristic imposes less overhead on the execution of the input workflow than the aggressive heuristic, but the potential for savings due to reuse is also lower. We quantify this tradeoff in the experiments in Section 7.

The technique we use to generate candidate sub-jobs for an input MapReduce job is as follows. We parse the physical plan of the input MapReduce job starting from its *Load* operators. For every parsed physical operator, we check if the heuristic that we are using requires us to generate a sub-job for this operator. If so, we inject a new *Store* operator after the parsed physical operator if the parsed operator is not already a *Store*. However, to include this *Store* operator in the physical plan of the input MapReduce job we need to also insert an operator that branches the output into two, similar to a Unix `tee` command. An example of this branching operator is the *Split* operator in Pig [11]. The output of the operator for which a sub-job is being generated is pipelined into the newly inserted *Split* operator. One branch of the output of the *Split* operator is pipelined into successor operators in the MapReduce job, and the other branch is pipelined into the new *Store* operator. Figure 8 shows the physical plan for Q1 after injecting two *Store* operators after the *Project* operators in the plan.

The sub-job generation step is performed for every MapReduce job in the input workflow of jobs. After this, ReStore submits the workflow of jobs to the MapReduce system for execution. The outputs of all MapReduce jobs in the workflow and the outputs of the injected *Store* operators (the sub-jobs) are all stored in the distributed file system. Storing the outputs of candidate sub-jobs introduces overhead, hence the need for the conservative and aggressive heuristics to choose these sub-jobs.

When storing generated jobs in the repository, ReStore ensures that the order of the physical plans in the repository follows the two ordering rules presented at the end of Section 3. These rules result in a repository in which the plans are partially ordered. The ordering of the plans in the repository ensures that the first plan from the repository that matches an input MapReduce job is the best match for this job.

## 5. MANAGING THE RESTORE REPOSITORY

Keeping the output of all generated jobs and sub-jobs in the repository can be expensive in the long run because of the storage space required and the increasing number of plans to match with future workflows. Therefore, for a set of candidate jobs and sub-jobs generated for an input workflow of MapReduce jobs as described in Section 4, we need to decide which of the outputs of these jobs to keep in the repository. This decision is made *after* the workflow is executed, so it is possible to base it on accurate execution statistics. In addition, we also need to decide when to evict stored job outputs from the repository.

In this paper, we focus on creating reuse opportunities and studying the cost and benefit of reuse. Therefore, we store the outputs of all candidate jobs and sub-jobs in the repository. Nevertheless, we present in this section some guidelines that can be used to decide which of the generated job outputs to store in the repository and when to evict stored outputs. A job output that is kept in the repository needs to satisfy two properties: (1) replacing the job with a *Load* of the job output from the distributed file system can reduce the execution time of a workflow that contains this job, and (2) there are future workflows that can reuse the output of this job.

We can check these properties based on statistics that the MapReduce system collects during job execution. For each candidate job or sub-job, we store in the repository statistics about the size of the input and output data, and the average execution time of the mappers and reducers. These statistics can easily be collected by any MapReduce system. For example, Hadoop already collects these statistics as part of job execution. We also collect and store statistics about the most recent time each job output in the repository was reused by another MapReduce job. These statistics can be used to decide which candidate jobs or sub-jobs to keep in the repository, and which jobs to evict from the repository. We propose the following rules for making these decisions. Rules 1 and 2 check the first property above (jobs in the repository reduce execution times when they are reused), and Rules 3 and 4 check the second property (jobs in the repository are actually reused).

1. Keep a candidate job in the repository only if the size of its output data is smaller than the size of its input data. This condition focuses on reducing $T_{load}$ in Equation 2, and therefore the total execution time of the job $ET(Job_i)$.

2. Keep a candidate job in the repository only if Equation 1 tells us that there will be a reduction in execution time for workflows reusing this job. This depends on the $max_{i \in Y}\{T_{total}(Job_i)\}$ component in Equation 1.

3. Evict a job from the repository if it has not been reused within a window of time.

4. Evict a job from the repository if one or more of its inputs is deleted or modified.

## 6. RESTORE IMPLEMENTATION

We have implemented ReStore as an extension to Pig 0.8 [2]. We briefly describe the Pig dataflow system and how its compiler generates a workflow of MapReduce jobs for a given query. We then present an overview of the implementation of ReStore [10].

### 6.1 Overview of the Pig Query Compiler

Pig [2, 11] is a dataflow system that compiles SQL-like queries written in the Pig Latin [14] query language into workflows of MapReduce jobs that are executed on Hadoop. The main stages of compiling a Pig Latin query are as follows: (1) a parser syntactically checks the input query and transforms it into a logical plan, which is a directed acyclic graph (DAG) of logical operators, (2) a logical optimizer applies optimization rules to this logical plan, (3) a MapReduce compiler transforms the logical plan into a physical plan and then compiles it into a series of MapReduce jobs, which forms a workflow, (4) a MapReduce optimizer applies rules to reduce the number of MapReduce jobs in the workflow, and (5) a Hadoop job manager submits the jobs in a workflow to Hadoop for execution taking into account the dependencies between them.

The JobControlCompiler is a component of the Hadoop job manager of Pig. Its input is a workflow of MapReduce jobs, where each job is represented by its physical plan. The JobControlCompiler iterates though the input workflow and decides on the jobs that can be run concurrently. It then prepares these jobs to be executed in Hadoop from their physical plans. After every iteration performed by the JobControlCompiler, a set of prepared jobs is submitted to Hadoop for execution. After the execution of the Hadoop jobs finishes, a new iteration of the JobControlCompiler is invoked. The outputs of the MapReduce jobs that are used as input to other jobs in the workflow are temporarily stored in the distributed file system (HDFS). After the completion of executing all the MapReduce jobs in the workflow, these intermediate outputs are deleted.

## 6.2 Implementation of ReStore

ReStore extends the JobControlCompiler of Pig. The input of ReStore is a workflow of MapReduce jobs. In every iteration of ReStore over the workflow, jobs that depend on already executed jobs or depend on no other jobs of the workflow are selected for execution, just as in the regular JobControlCompiler. Every physical plan of these jobs passes though two stages: (1) matching with plans in the repository, and (2) generating candidate sub-jobs. MapReduce jobs are then prepared for the rewritten physical plans and are submitted to Hadoop for execution using the same techniques used by the JobControlCompiler. After executing a MapReduce job, statistics about this job execution are retrieved from Hadoop and stored in the repository to decide which job outputs to keep (Section 5). We also store information about the physical plans of stored jobs in the repository. We implement the repository as a table that contains in every record: (1) a physical plan of a MapReduce job, (2) the filename of the output of this job in HDFS, and (3) statistics about this job.

## 7. EXPERIMENTS

We conducted our experiments on a cluster of 15 servers (henceforth referred to as "nodes"). Each node has four Dual Core AMD Opteron 275 processors running at 2.2 GHz and 8GB of memory. The nodes run SuSE Linux 10.1. Each node has a 65GB SCSI disk, and the HDFS file system is created on these disks. The Hadoop cluster is configured to have the Hadoop TaskTracker and NameNode running on one dedicated node and each of the remaining 14 nodes running a TaskTracker and a DataNode. Each TaskTracker can run a maximum of 4 mappers and a maximum of 2 reducers simultaneously.

We use the PigMix [4] benchmark in our experiments. We generate the data using the PigMix data generator to create two instances of the benchmark data: (1) an instance where the `page_views` table has 10 million rows and a size of approximately 15GB in HDFS before its 3-way replication, and (2) an instance where this table has 100 million rows and a size of approximately 150GB before 3-way replication. The generated instances include other tables, but these tables are much smaller than the `page_views` table, so we refer to these two instances as the 15GB and 150GB instances. In our experiments we use a subset of the benchmark consisting of queries L2–L8 and L11. These queries test a wide range of features and operators (described in detail in [4]) that include *Join*, *Group*, *CoGroup*, *Filter*, *Distinct*, and *Union*. We have excluded L1, L9, and L10 from our experiments because they test features that are not relevant to result reuse. To illustrate the effectiveness of reusing the output of whole jobs, we also created a synthetic workload that is based on queries L3 and L11 of PigMix, which are each translated by Pig into a workflow of multiple MapReduce jobs (more details in Section 7.1). We also use a completely synthetic workload that is not based on PigMix in Section 7.5.

Each reported result is based on the average execution time of three runs as measured by Hadoop. The workflows that we evaluate are: (1) unmodified workflows as generated by Pig, (2) workflows after injecting extra *Store* operators by ReStore to materialize sub-jobs, and (3) workflows after being rewritten by ReStore to reuse job outputs in the repository. In addition to execution time, we use as a performance metric the *speedup* of a query, which is the execution time of the query when no data reuse is considered divided by the execution time of the query when it is rewritten to reuse job outputs in the repository. We also use as a metric the *overhead* of adding *Store* operators to a query, which is the execution time of the query when extra *Store* operators are injected into the physical plans of its MapReduce jobs divided by the execution time of the query when the physical plans are unchanged.

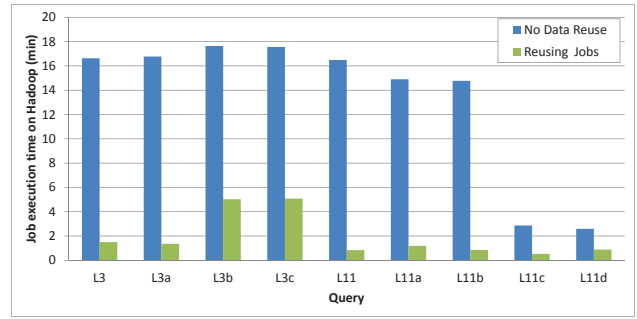

Figure 9: The effect of reusing whole job outputs for data size 150GB.

## 7.1 Reusing the Output of Whole Jobs

In this section, we demonstrate the effectiveness of reusing the output of whole jobs for rewriting the MapReduce workflows of future queries. In Figure 9, we compare the following: (1) the execution time of queries without reusing outputs from prior executions, and (2) the execution time when reusing outputs of whole jobs that have been stored during previous executions of the same query by Hadoop. In the latter case, we are assuming that all outputs of jobs that can be reused by a given query are available in the ReStore repository. Therefore, the execution times of the PigMix queries reported in Figure 9 are the best that can be achieved when using the current implementation of ReStore and reusing the output of prior executed jobs.

For this experiment we use queries L3 and L11 of the PigMix benchmark. These two queries are translated by Pig into workflows of more than 1 MapReduce job (2 jobs for L3 and 3 for L11) making it possible to reuse the output of whole jobs. The workflow of query L11 contains 3 jobs, where one job depends on the other two. We generated variants of the queries to increase the diversity of the workload. Query L3 does grouping and aggregation, and we changed the aggregation function to generate different L3 variants. Query L11 combines two data sets using a union operation, and we changed the data sets that are combined. Figure 9 shows that the average speedup due to job reuse is 9.8. The overhead in this case is 0% since no extra *Store* operators are inserted in the physical plan. Thus we can see that ReStore can be highly beneficial.

## 7.2 Reusing the Output of Sub-Jobs

In this section, we illustrate the benefit of reusing the outputs of sub-jobs generated by ReStore. For every PigMix query, we report the following: (1) the execution time of the query without reusing outputs from prior execution, (2) the execution time of the query without reusing outputs from prior execution, but when injecting *Store* operators in the physical plans of the executed MapReduce jobs to materialize the outputs of sub-jobs that are selected by the Aggressive Heuristic described in Section 4 (we compare the two heuristics proposed in Section 4 in the next section), and (3) the execution time when reusing outputs of sub-jobs that have been stored in the repository during previous executions of the same query by Hadoop. The execution times reported for the last case are measured when all the sub-jobs that are generated by ReStore are available in the repository. Figure 10 shows the results of this comparison for data size 150GB.

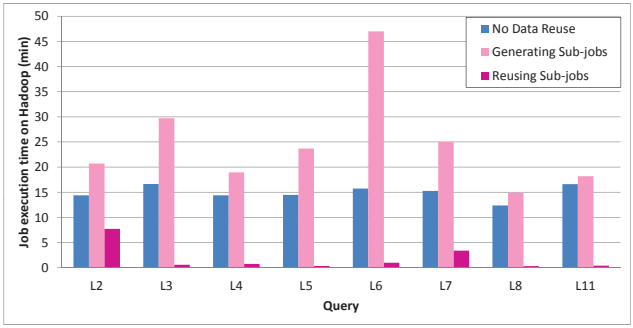

Figure 10: The effect of reusing sub-job outputs for data size 150GB.

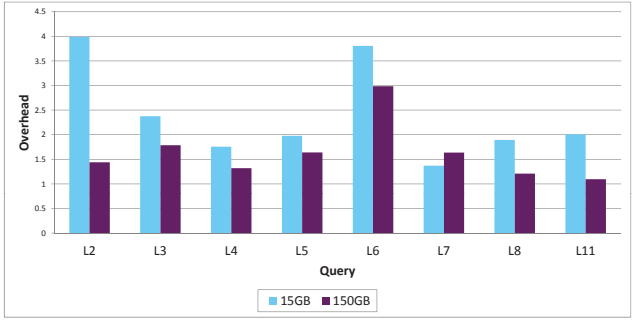

Figure 11: Overhead for data data sizes 15GB and 150GB.

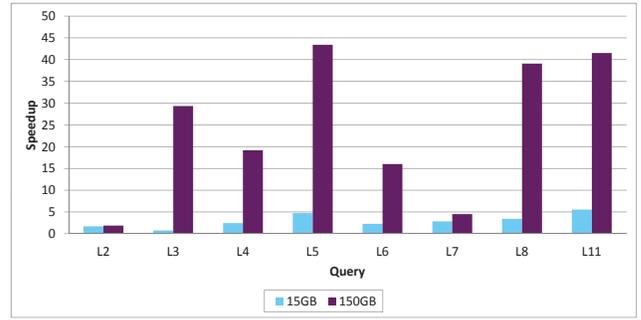

Figure 12: Speedup for data data sizes 15GB and 150GB.

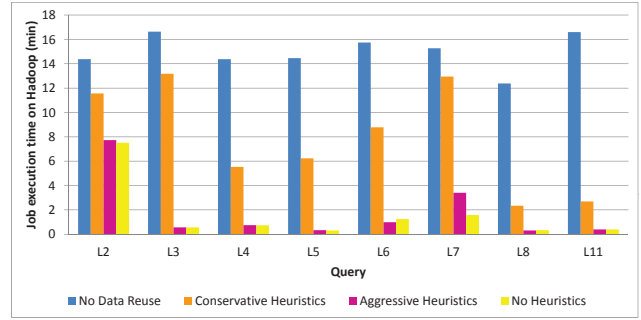

Figure 13: Execution time when reusing sub-jobs chosen by different heuristics (150GB).

Figure 10 shows that the average speedup achieved by reusing sub-job outputs is 24.4. There is an overhead incurred due to injecting extra *Store* operators for materializing sub-job outputs. This overhead is seen as an increase in the execution time of queries (1.6 on average). The overhead for L6 is high since a *Store* operator is injected in the reducer after an expensive *Group* operator whose output size is large (5.5 GB). The number of reducers is small and storing large data in them increases the execution time significantly. This is the nature of the Aggressive Heuristic: the speedup due to result reuse is very high, but the overhead can potentially be high too. In this experiment, using ReStore was beneficial if the output of a sub-job is reused even only once in the future.

Figure 11 shows the overhead incurred when adding *Store* operators to the physical plans of MapReduce jobs of the PigMix queries for data sizes 15GB and 150GB. For most PigMix queries, the overhead incurred when executing on the 15GB data set is higher than that when executing on the 150GB data set. The average overhead is 2.4 for the 15GB data set and 1.6 for the 150GB data set. For small data sizes, a slight increase in $T_{store}$ can have a big effect on $ET(Job_n)$ because the time spent in loading the data and executing the operators is small (Equation 2). However, increasing $T_{store}$ will appear less significant for large data sizes since $T_{load}$, $\sum_i ET(OP_i)$, and $T_{sort}$ are high.

Figure 12 shows the speedup achieved by reusing sub-job outputs for data sizes 15GB and 150GB. The speedup achieved by most PigMix queries is higher for the larger data size. While the average speedup is 3.0 for the 15GB case, it increases to 24.4 for the 150GB case. In Equation 2, replacing $T_{load}$ and part of $\sum_i ET(OP_i)$ with $T_{load}^R$, which is the time needed to load sub-job outputs, has a larger effect on $ET(Job_n)$ for the larger data size because $T_{load}$ is the most expensive operation in this case. This experiment demonstrates that reusing the output of sub-jobs is highly effective, and it is more beneficial for larger data sizes.

### 7.3 Comparing the Heuristics for Generating Candidate Sub-Jobs

Section 4 describes two heuristics for choosing the physical operators in an input MapReduce job whose outputs to materialize as sub-jobs, the Conservative Heuristic and the Aggressive Heuristic. If we do not use these heuristics, we inject a *Store* operator after each physical operator in the input MapReduce job. We call this the No Heuristic case. In this experiment, we compare the performance of No Heuristic (NH), the Conservative Heuristic ($H_C$), and the Aggressive Heuristic ($H_A$). In all of these cases, the outputs of *Store* operators added to the MapReduce jobs are stored in the distributed file system, which adds overhead to the execution of the MapReduce workflows and occupies storage in the distributed file system. We expect NH to have the highest overhead and require the most storage, followed by $H_A$, and then $H_C$. Our goal in this experiment is to verify this and quantify the difference between the three strategies. A related question is whether the extra overhead and storage requirement of a policy such as NH results in more reuse opportunities and higher speedups.

| Q | I/P (GB) | $H_C$ (GB) | $H_A$ (GB) | NH (GB) | O/P |
|---|---|---|---|---|---|
| L2 | 150.6 | 3.1 | 3.1 | 6.7 | 1.1 MB |
| L3 | 150.7 | 3.2 | 8.2 | 22.1 | 62.9 MB |
| L4 | 150.6 | 2 | 2.8 | 10.8 | 34.2 MB |
| L5 | 150.7 | 1.8 | 4.6 | 7.4 | 2 B |
| L6 | 150.6 | 3.7 | 10.1 | 24.3 | 92.7 MB |
| L7 | 150.6 | 2.2 | 5.4 | 5.4 | 1.5 MB |
| L8 | 150.6 | 3.3 | 3.3 | 11.4 | 27 B |
| L11 | 173.6 | 2.6 | 2.7 | 2.8 | 1.6 GB |

Table 1: Total size of input data loaded by different queries, output of *Store* operators added by different heuristics, and final query output.

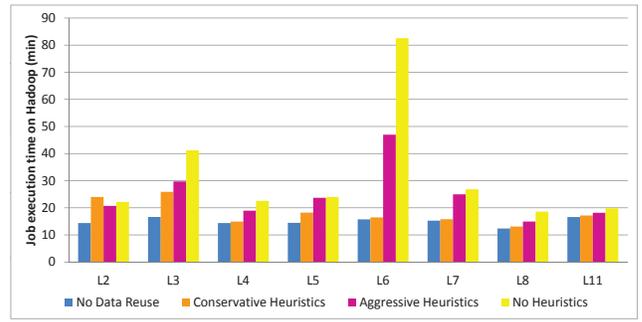

Figure 14: Execution time with the *Store* operators chosen by different heuristics to materialize sub-job outputs (150GB).

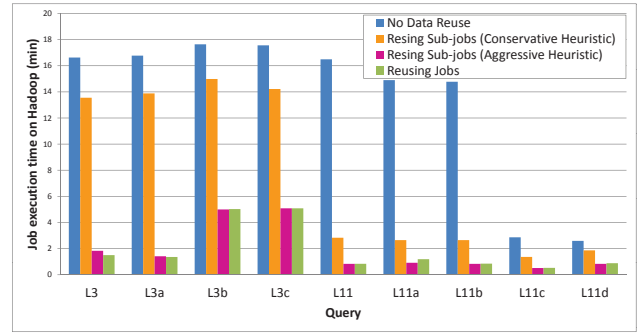

Figure 15: Reusing whole jobs and sub-jobs for data size 150GB.

Figure 13 compares the execution time of PigMix queries without result reuse and when reusing the sub-jobs stored by each of NH, $H_C$, and $H_A$. The figure verifies our expectation that $H_A$ provides more reuse opportunities and hence higher benefit than $H_C$. A very encouraging result from the figure is that $H_A$ has the same performance as NH. This is because $H_A$ stores the output of the main set of operators that increase the execution time of a MapReduce job. The additional sub-job outputs stored by NH provide no benefit. Also, the fact that $H_C$ stores fewer sub-jobs leads to lower benefit from sub-job reuse. Therefore, we use $H_A$ as our default heuristic, as we have seen in the previous section (Figures 10–12).

The drawbacks of using $H_A$ are the extra storage required in the distributed file system and the increase in the execution time of MapReduce jobs due to the extra *Store* operators. Table 1 shows for each PigMix query the size of the loaded data, the size of the data produced by the extra *Store* operators under $H_C$, $H_A$, and NH, and the size of the final query output. The extra data stored by the $H_A$ is always much less than NH and usually close to $H_C$. This supports our choice of the Aggressive Heuristic as the default heuristic. There are, however, some cases where $H_A$ stores much more data than $H_C$, such as L6. Figure 14 shows the execution times of the queries with the extra *Store* operators under the three heuristics plus the time with no data reuse (i.e., no storage of sub-jobs). The conclusions drawn previously are corroborated by this experiment: $H_A$ is always better than NH and usually only slightly worse than $H_C$, but there are cases like L6 where $H_A$ is much worse than $H_C$.

Thus, we conclude that using the Aggressive Heuristic is better overall from the point of view of overhead and potential speedup, but it does incur some risk. If we desire to reduce this risk, we can use the Conservative Heuristic which reduces overhead but sacrifices some speedup.

### 7.4 Reusing Sub-Jobs vs. Whole Jobs

Up to this point, we have seen the results of different types of reuse in ReStore: reusing the results of whole jobs, reusing the results of sub-jobs chosen by the Conservative Heuristic $H_C$, and reusing the results of sub-jobs chosen by the Aggressive Heuristic $H_A$. In this experiment, we summarize the difference between these three types of reuse. Queries L3 and L11 and their variants are the queries that we used for evaluating the reuse of whole intermediate jobs. In Figure 15 we plot the execution time of these queries with no reuse and with reusing the results of whole intermediate MapReduce jobs. The figure also shows the execution time of these queries when reusing sub-jobs chosen using $H_C$ and $H_A$. We make two observations about this figure. First, we see that all types of reuse in ReStore are beneficial, although to varying degrees. Second, we see that the maximum benefit is obtained when reusing whole jobs or sub-jobs chosen by $H_A$, and that the difference between these two cases is minimal.

Reusing whole jobs is expected to yield the maximum benefit and it also has the advantage that it does not incur any overhead since the job outputs are already stored in the distributed file system. However, it is not always possible to reuse whole jobs, which is why ReStore generates sub-jobs. Furthermore, sub-jobs are less specific than whole jobs so they can be useful for more queries. In Figure 15, the sub-jobs chosen by $H_A$ are indeed less specific and they do not include all the physical operators that are present in the whole jobs. Thus, reusing sub-jobs chosen by $H_A$ in-

595

| Field name | Cardinality | % Selected Data |
|---|---|---|
| *field6* | 200 | 0.5% |
| *field7* | 100 | 1% |
| *field8* | 20 | 5% |
| *field9* | 10 | 10% |
| *field10* | 5 | 20% |
| *field11* | 2 | 50% |
| *field12* | 1.6 | 60% |

Table 2: Fields of the generated synthetic data set.

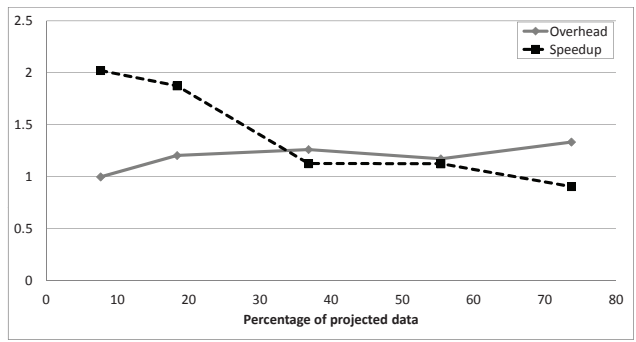

Figure 16: The overhead and speedup of different jobs with *Project* operators.

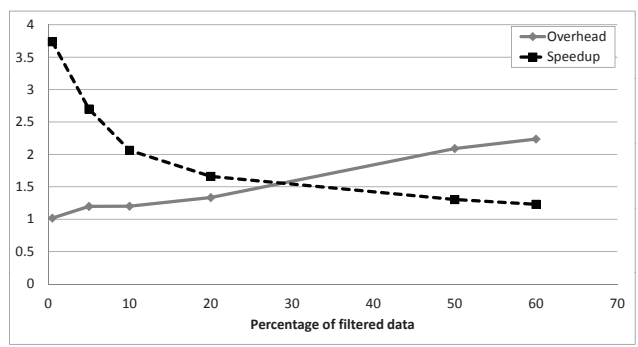

Figure 17: The overhead and speedup of different jobs with *Filter* operators.

curs some extra work when compared to reusing whole jobs. This extra work is needed to apply the physical operators in the whole jobs that are missing from the sub-jobs. The fact that reusing sub-jobs chosen by $H_A$ is as good as reusing whole jobs tells us that the Aggressive Heuristic is effective at capturing the most expensive parts of a MapReduce job while avoiding cheap parts that can be applied easily.

Thus, we see that reusing whole jobs and sub-jobs are both important. Reusing whole jobs is cheap and effective, but it is not always possible. However, it is always possible to create reuse opportunities by generating sub-jobs, and $H_A$ does so quite effectively, albeit at a cost. If this cost is a concern, a reasonable benefit can still be derived by using the lower cost $H_C$.

## 7.5 Effect of Data Reduction

One of the benefits of storing sub-jobs, particularly the output of *Filter* and *Project* operators, is that future workflows that reuse these sub-jobs will read less data as compared to reading the original data set. As the amount of data eliminated by the *Filter* or *Project* operator increases, the overhead of storing the sub-job decreases and the benefit of reusing it increases. In this experiment, we aim to demonstrate this effect.

For this experiment, we generated a data file that contains 200 million rows using the same data generator used to generate the PigMix data set. The size of the generated data is 40GB. The generated data file has 12 synthetic fields, *field1* through *field12*. Fields *field1* through *field5* are random strings of length 20 characters each, used to study the *Project* operator. Fields *field6* through *field12* are integers that are used to study the *Filter* operator, and each of them has a different cardinality (number of distinct values) so that when we apply an equality predicate on each field we select a different percentage of rows from the input table.

The cardinality and the percentage of rows selected by an equality predicate for each field are shown in Table 2.

We first study the effect of data reduction in *Project* operators. For this experiment we created a workload of 5 queries on the synthetic data set that follow the template QP (below). We vary the number of fields selected by the *Project* operator. At a minimum we select *field1*, and at a maximum we select *field1* through *field5*. Our data set is designed such that when one field is selected by the *Project* operator, the size of the output of this operator is around 18% of the original data size and when all five fields are selected the output is around 74% of the original data.

**Query Template QP**
```
A = load '$synth_data' as (field1, ..., field12);
B =  foreach A generate  field1, ...;
C = group B by (field1, ...);
D = foreach C generate COUNT($1);
store D into '$out';
```

We compare the execution time of QP without data reuse, when we inject an extra *Store* operator in the physical plan of the MapReduce job after the *Project* operator, and when we reuse the output of this *Store* operator. Figure 16 shows the overhead of the extra *Store* operator and the speedup achieved by reusing the output of this operator as we vary the number of projected fields from 1 to 5. The figure shows that as the amount of data reduction due to projection decreases, the overhead increases and the speedup decreases. In this experiment, if the *Project* operator reduces the size of the input data by more than half, there will be a net benefit if this stored data is reused at least once.

Next, we turn our attention to *Filter* operators. For this we use query template QF (below), in which we apply an equality predicate on one of the fields *field6* to *filed12*. The data reduction for each field varies as shown in Table 2.

**Query Template QF**
```
A = load '$synth_data' as (field1, ..., field12);
B = filter A by $fieldi = $val ;
C = group B by field1;
D = foreach C generate COUNT($1);
store D into '$out';
```

Figure 17 shows the overhead of injecting an extra *Store* operator after the *Filter* and the speedup due to reusing the output of this operator for the six different instantiations of QF. As before, when the amount of data reduction decreases, the overhead increases and the speedup decreases.

## 8. RELATED WORK

Distributed analysis of large data through the MapReduce model was introduced by Google [9]. Since then, several implementations of MapReduce have appeared, the most prominent being Hadoop [1]. Dataflow language processors such as Pig [11], Hive [15], and Jaql [8] provide an easy way to write SQL-like queries that are translated into workflows of MapReduce jobs, thereby enabling users to easily express more complex analysis tasks.

There has been work on several types of optimizations for MapReduce, but in this paper we focus on one specific type of optimization, namely sharing computation between different MapReduce jobs. This type of optimization has been explored before in [5]. In that work, a study is presented of how jobs submitted to a MapReduce system can be scheduled to benefit from sharing *scans over a common set of files*. New policies for scheduling MapReduce jobs are introduced with the goal of maximizing the likelihood of sharing scans. That work differs from ours in that it exploits one specific type of sharing opportunity, and that the sharing happens between concurrently running MapReduce jobs. In contrast, our paper can exploit different types of sharing opportunities, not just of scans (which correspond to *Load* operators) but of other more complex physical plans. Moreover, our paper enables sharing between jobs that are executed at different times by storing and reusing job outputs.

Another work that attempts to share computation in MapReduce is MRShare [13]. MRShare finds sharing opportunities among queries that are submitted in the same batch to the MapReduce data analysis platform. The main goal of MRShare is to avoid redundant work by combining the execution of operators from different queries. In [13], a cost model is also proposed to find the optimal plan for merging a group of queries appearing in the same batch. As with [5], that work differs from ours in that we focus on reducing work for individual jobs by leveraging previously stored job outputs, and not for batches of jobs.

Materializing the results of executed queries to be reused by future queries has been extensively studied in the context of materialized views for relational databases [12]. Most commercial database systems now include a physical design advisor that automatically recommends materialized views given a sample workload of queries (e.g. [6]). In this paper, we focus on creating reuse opportunities that have the same essence as these previous works, but the details are fundamentally different because of the distinct challenges introduced by MapReduce, namely the massive data sizes and the procedural nature of the query languages.

## 9. CONCLUSION

The MapReduce model has become widely accepted for analyzing large data sets. In many cases, users express complex analysis tasks not directly with MapReduce but rather with higher-level SQL-like query languages that get translated into workflows of MapReduce jobs. It is important to improve the performance of these workflows given how widely adopted they are. When executing a workflow of MapReduce jobs, intermediate outputs of jobs are stored in the distributed file system to be read by subsequent jobs. This behavior creates an opportunity for performance optimization: instead of deleting the output of intermediate jobs, store this output for reuse in answering future workflows that are executed by the system and that perform the same computation as the job whose output is stored.

In this paper, we present ReStore, a system that reuses intermediate outputs of MapReduce jobs in a workflow to speed up future workflows executed in the system. In addition to reusing the outputs of whole MapReduce jobs, ReStore can also create more reuse opportunities by storing the output of some physical query execution operators that form part of a MapReduce job. We have implemented ReStore as an extension to the Pig dataflow system. Our experiments with this implementation show that ReStore can achieve speedup ups to an order of magnitude, and the overhead it incurs is not prohibitive

**Acknowledgements** This work was supported by the Natural Sciences and Engineering Research Council of Canada (NSERC) through the Business Intelligence Network strategic networks grant.